# Observation of giant bandgap renormalization and excitonic effects in a monolayer transition metal dichalcogenide semiconductor

Miguel M. Ugeda[1,†,*], Aaron J. Bradley[1,†], Su-Fei Shi[1,†], Felipe H. da Jornada[1,2], Yi Zhang[3,4], Diana Y. Qiu[1,2], Sung-Kwan Mo[3], Zahid Hussain[3], Zhi-Xun Shen[4,5], Feng Wang[1,2,6,*], Steven G. Louie[1,2], Michael F. Crommie[1,2,6,*].

[1] *Department of Physics, University of California at Berkeley, Berkeley, California 94720, USA.*

[2] *Materials Sciences Division, Lawrence Berkeley National Laboratory, Berkeley, California 94720, USA.*

[3] *Advanced Light Source, Lawrence Berkeley National Laboratory, Berkeley, CA 94720, USA.*

[4] *Stanford Institute for Materials and Energy Sciences, SLAC National Accelerator Laboratory, Menlo Park, CA 94025, USA.*

[5] *Geballe Laboratory for Advanced Materials, Departments of Physics and Applied Physics, Stanford University, Stanford, CA 94305, USA.*

[6] *Kavli Energy NanoSciences Institute at the University of California Berkeley and the Lawrence*
*Berkeley National Laboratory, Berkeley, CA 94720, USA.*

*Corresponding authors: mmugeda@berkeley.edu (MMU), fengwang76@berkeley.edu (FW) and crommie@berkeley.edu (MFC).

† These authors contributed equally to this work.

**Two-dimensional (2D) transition metal dichalcogenides (TMDs) exhibit novel electrical and optical properties and are emerging as a new platform for exploring 2D semiconductor physics[1-9]. Reduced screening in 2D results in dramatically enhanced electron-electron interactions, which have been predicted to generate giant bandgap renormalization and excitonic effects[10-13]. Currently, however, there is little direct experimental confirmation of such many-body effects in these materials. Here we present an experimental observation of extraordinarily large exciton binding energy in a 2D semiconducting TMD. We accomplished this by determining the single-particle electronic bandgap of single-layer $MoSe_2$ via scanning tunneling spectroscopy (STS), as well as the two-particle exciton transition energy via photoluminescence spectroscopy (PL). These quantities yield an exciton binding energy of 0.55 eV for monolayer $MoSe_2$, a value that is orders of magnitude larger than what is seen in conventional 3D semiconductors. This finding is corroborated by our ab initio GW and Bethe Salpeter equation calculations[14, 15] which include electron correlation effects. The renormalized bandgap and large exciton binding observed here will have a profound impact on electronic and optoelectronic device technologies based on single-layer semiconducting TMDs.**

The remarkable properties of atomically-thin semiconducting TMD layers include an indirect-to-direct bandgap crossover[1, 2, 9], field-induced transport with high on-off ratios[16], valley selective circular dichroism[3-6], and strong photovoltaic response[17, 18]. Fundamental understanding of the electron/hole quasiparticle band structure and many-body interactions in 2D TMDs, however, is still lacking. Enhanced Coulomb interactions due to low-dimensional effects are expected to increase the quasiparticle bandgap as well as to cause electron-hole pairs to form more strongly bound excitons[10-13]. Untangling such many-body effects in single-layer TMDs requires measurement of both the electronic bandgap and the optical bandgap, the most fundamental parameters for transport and optoelectronics, respectively. The electronic bandgap ($E_g$) characterizes single-particle (or quasiparticle) excitations and is defined by the sum of the energies needed to separately tunnel an electron and a hole into monolayer $MoSe_2$. The optical bandgap ($E_{opt}$), on the other hand, describes the energy required to create an exciton, a correlated two-particle electron-hole pair, via optical absorption. The difference in these energies ($E_g - E_{opt}$) directly yields the exciton binding energy ($E_b$) (Fig. 2a). Here we provide evidence for Coulomb driven quasiparticle bandgap renormalization and unusually strong exciton stability in 2D TMD through direct determination of both $E_g$ and $E_{opt}$ via STS and PL spectroscopy, respectively.

STS and PL measurements were carried out on the same high-quality sub-monolayer $MoSe_2$ films grown on epitaxial bilayer graphene (BLG) on a 6H-SiC(0001) substrate. Because the $MoSe_2$ surface coverage for our sample was ~ 0.8 ML, we were able to simultaneously image the $MoSe_2$ monolayer and the underlying graphene substrate using scanning tunneling microscopy (STM). Figs. 1b and 1d show atomically-resolved STM images taken of a $MoSe_2$ monolayer region and a BLG substrate region, respectively. The BLG substrate (Fig. 1d) shows typical hexagonal atomic contrast due to the Bernal AB stacking overlaid with the (6√3 x 6√3) SiC reconstruction[19]. The high-resolution STM image acquired on MoSe$_2$ (Fig. 1b) shows a honeycomb atomic lattice with one sublattice brighter than the other. Both sublattices show a unit cell of 3.3 Å, corresponding to the interatomic spacing in both the basal Se and Mo planes of MoSe$_2$ (see SI). The larger-scale MoSe$_2$ image (Fig. 1c) shows only one sublattice due to a slightly lower (and more typical) spatial resolution. The atomic registry of the MoSe$_2$ lattice was often oriented precisely with the graphene lattice (yellow arrows in Figs. 1b and d) but also frequently showed a slight rotation angle (dashed purple lines in Fig. 1c). An additional periodic superlattice was always visible in the STM images (Figs. 1b and 1c). This superlattice is explained by the fact that when the MoSe$_2$ and graphene lattices are overlaid, four graphene unit cells accommodate three unit cells of MoSe$_2$, thus forming a quasi-commensurate 9.87 Å x 9.87 Å superstructure (a moiré pattern) as observed in the STM images of MoSe$_2$. Rotationally misaligned MoSe$_2$ domains show slightly smaller incommensurate moiré patterns (see SI).

We experimentally investigated both the electronic structure and the optical transitions in monolayer MoSe$_2$/BLG by combining STS and PL spectroscopy. Fig. 2b shows a typical STM dI/dV spectrum acquired on monolayer MoSe$_2$/BLG. The observed electronic structure is dominated by a large electronic bandgap surrounded by features labeled V$_{1-4}$ in the valence band (VB) and C$_1$ in the conduction band (CB). The MoSe$_2$ band edges are best determined by taking the logarithm of dI/dV, as shown in Fig. 2d. There the VB maximum (VBM) for monolayer MoSe$_2$ is seen to be located at -1.55 ± 0.03 V and the CB minimum (CBM) at 0.63 ± 0.02 V. The relative position of E$_F$ (V$_{bias}$ = 0 V) with respect to the band edges reveals n-type doping for our samples, although with a very low carrier concentration. We tentatively attribute the n-doping of our $MoSe_2$ samples to intrinsic point defects such as vacancies and/or lattice antisites, which have been found to be responsible for n-doping in similar materials[20]. Our STS measurements yield a value for the single-particle electronic bandgap of $E_g = E_{CBM} - E_{VBM} = 2.18$ eV ± 0.04 eV. The uncertainty in Eg arises from both lateral positional variations as well as tip-induced band bending (TIBB) (see SI). TIBB can arise due to poor screening of electric fields at a semiconducting surface[21]. We are able to rule out TIBB as a significant source of error in our $E_g$ measurement via two methods: (i) tip-sample distance variation (which allows us to sample different electric field values) and (ii) comparison of our spectra with our angle-resolved photoemission spectroscopy (ARPES). We find that the $V_1$ feature, for example, arises from the spin-split valence band at the K-point that is seen for single layer semiconducting TMDs (see SI for a more detailed discussion).

Measurement of the optical bandgap ($E_{opt}$), i.e. the electron-hole excitation energy, was performed using PL spectroscopy. PL from single layer $MoSe_2$ is partially quenched by the graphene underneath, but the PL signal is still clearly observable. A representative set of data taken at two different temperatures is shown in Fig. 3. At room temperature, the PL spectra from monolayer $MoSe_2$ show a clear Lorentzian shape centered at 1.55 ± 0.01 eV. At 77 K, the peak position of the PL is shifted to 1.63 ± 0.01 eV. This decrease of photon energy at high temperature has been observed previously[22] and is attributed to thermal reduction of the $MoSe_2$ electronic bandgap. Since previous studies[22] have shown that PL does not shift significantly when temperature drops below 77 K, we determine the low-temperature $MoSe_2$ optical bandgap to be $E_{opt} = 1.63 ± 0.01$ eV (uncertainty here arises from spatial variations and temperature dependence).

The optical bandgap of $E_{opt}$ = 1.63 eV differs from the electronic bandgap $E_g$ = 2.18 eV by the amount 0.55 ± 0.04 eV, which corresponds to the binding energy of the electron-hole excitation (i.e., the exciton). This large value for the $MoSe_2$ excitonic binding energy is nearly two orders of magnitude higher than the binding energy seen in conventional semiconductors such as Si or Ge, and has significant implications for the optoelectronic properties of TMD materials even at room temperature.

In order to better understand and interpret our experimental findings, we performed *ab initio* GW[14] and GW plus Bethe-Salpeter equation (GW-BSE) calculations[15, 23] on $MoSe_2$ using the BerkeleyGW package[23]. This allowed simulation of monolayer $MoSe_2$/BLG electronic structure and optical transitions. Similar to previous calculations on related $MoS_2$[13], we found it necessary to employ a large energy cutoff of 38 Ry and to use a large number (10000) of unoccupied states to obtain well-converged quasiparticle energies (see SI). Within this approach we find that $MoSe_2$ monolayers are direct gap semiconductors at both the DFT and GW levels of theory. For quantitative comparison between experiment and theory we had to take into account the screening of the $MoSe_2$ layer by the BLG underneath. This is expected to decrease both the quasiparticle bandgap and the exciton binding energy. Since it is too computationally demanding to perform a highly converged calculation for the explicit $MoSe_2$/BLG supercell system, we developed a novel method to incorporate substrate screening. We first separated the screening into intrinsic and substrate contributions[24]. We then fully took into account the in-plane substrate long wavelength screening and the full perpendicular component of the screening while neglecting in-plane local fields produced by the substrate (this is justified by the in-plane delocalization of the $\pi$ orbitals in graphene). Within this approach we determined a quasiparticle bandgap of 2.13 ± 0.10 eV for MoSe$_2$ deposited on BLG, in good agreement with the experiment. The theoretical uncertainty arises primarily due to the GW approximation of the electronic self-energy (see SI).

Combining this theoretical approach with the GW-BSE method allowed us to calculate the optical spectrum and the exciton binding energy of a MoSe$_2$ monolayer both with and without substrate screening. We find that the BLG substrate decreases the exciton binding energy from 0.65 ± 0.05 eV to 0.52 ± 0.05 eV, in good agreement with the experimental value of 0.55 ± 0.04 eV. Similarly, the calculated optical gap for MoSe$_2$ on BLG is found to be 1.61 ± 0.11 eV, in good agreement with the measured optical gap of 1.63 ± 0.01 eV. The calculated electronic bandgap, optical gap, and exciton binding energies are all graphically compared with the experimental results in the energy level diagrams of Figs. 4a-c. In Fig. 4d, we compare our calculated optical absorption spectrum with the experimental PL data. The comparison here is quite good (note that PL typically only measures the lowest absorption peak since electron-hole excitations usually relax quicker than the fluorescence lifetime). Fig. 4e shows the predicted spatial dependence of the exciton state explored in this study.

Our calculations and measurements were performed on undoped or nearly undoped samples, where many-electron interactions are prominent. For more heavily doped TMD samples (or, similarly, samples with more heavy substrate screening) we expect the self-energy correction to the quasiparticle bandgap and the excitonic effects to be reduced by free carrier screening, which should bring the quasiparticle bandgap closer to the optical one[25, 26]. This has indeed been observed in previous ARPES measurements on highly doped MoSe$_2$ [9].

In conclusion, the strong excitonic binding energy observed here should play a significant role in the room-temperature performance of optoelectronic nanodevices composed of single-layer semiconducting TMDs as well as more complex layered heterostructures. For example, the large difference between $E_g$ and $E_{opt}$ (0.55 eV) will affect both the design and evaluation of photodetectors and photovoltaics (which exploit the dissociation of excitons). The large exciton binding energy observed here also highlights the importance of many-body effects in atomically thin 2D layers.

**Methods:**

The measurements were carried out on high-quality sub-monolayer $MoSe_2$ films grown by molecular beam epitaxy on epitaxial BLG on a 6H-SiC(0001) substrate (Fig. 1). The structural quality of the samples and the $MoSe_2$ coverage were characterized by in-situ reflection high electron diffraction (RHEED), low energy electron diffraction (LEED), core level spectroscopy, and Raman spectroscopy (see Supplementary information). Scanning tunneling microscopy (STM) imaging and STS experiments were performed in an ultra-high vacuum (UHV) system equipped with a home-built STM operated at T = 5 K. STM differential conductance (dI/dV) spectra were measured at 5 K by using standard lock-in techniques. In order to avoid tip artifacts, the STM tip was always calibrated by measuring reference spectra on the BLG substrate[27, 28] (see SI). PL experiments were performed in high-vacuum using continuous wave excitation centered at 532 nm with a power of 500 μW and a focused spot size of 2 μm. Different spots across the sample consistently showed similar PL and STS spectra. Both STM/STS and PL measurements were performed consecutively on the same samples.

**Acknowledgments:**

Research supported by Office of Basic Energy Sciences, Department of Energy Early Career Award No. DE-SC0003949 (optical measurements), and by the sp$^2$ Program (STM instrumentation development and operation), the Theory Program (GW-BSE calculations), and the SciDAC Program on Excited State Phenomena in Energy Materials (algorithms and codes) which are funded by the U. S. Department of Energy, Office of Basic Energy Sciences and of Advanced Scientific Computing Research, under Contract No. DE-AC02-05CH11231. Support also provided by National Science Foundation award DMR-1206512 (image analysis) and National Science Foundation Award No. DMR10-1006184 (substrate screening theory and calculations). Computational resources have been provided by the NSF through XSEDE resources at NICS and DOE at NERSC. A.J.B. was supported by the Department of Defense (DoD) through the National Defense Science & Engineering Graduate Fellowship (NDSEG) Program. D. Y. Q. acknowledges support from NSF Graduate Research Fellowship Grant No. DGE 1106400 and SGL acknowledges support of a Simons Foundation Fellowship in Theoretical Physics. STM/STS data were analyzed and rendered using WSxM software[29]. S.-F. S and F. W. acknowledge X. Hong and J. Kim for technical help.

**Author contributions:**

M.M.U., A.J.B., S-.F.S., F.W., and M.F.C. conceived the work and designed the research strategy. M.M.U. and A.J.B. measured and analyzed the STM/STS data. S-.F.S. carried

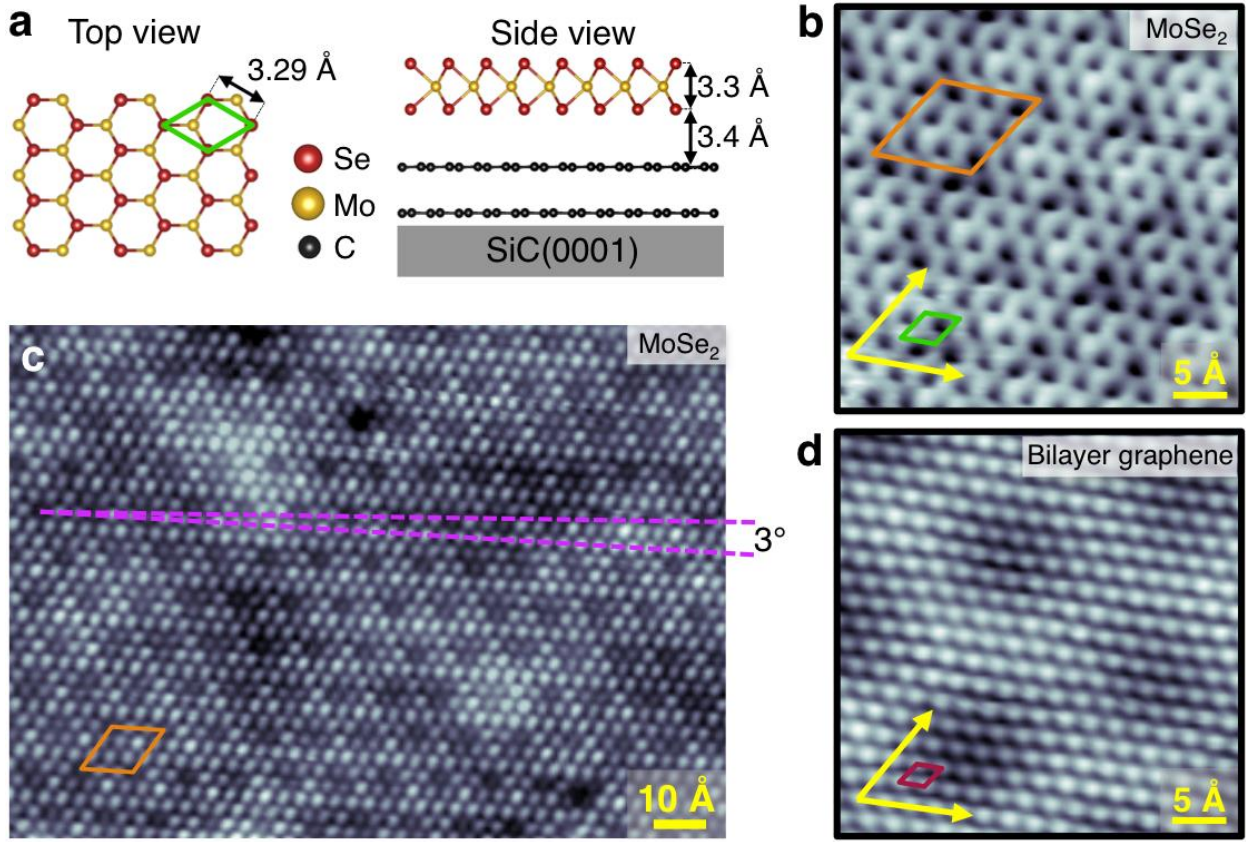

**Figure 1. Morphology of monolayer MoSe$_2$ on bilayer graphene (BLG). a**, Top and side view sketches of monolayer MoSe$_2$, including the substrate. **b**, High-resolution STM image of MoSe$_2$ ($V_s$ = - 1.53 V, $I_t$ = 3000 pA, T = 5 K). **c**, STM image (typical resolution) of monolayer MoSe$_2$ showing 9.7 Å x 9.7 Å moiré pattern with an angle of 3° between the moiré pattern and the MoSe$_2$ lattice ($V_s$ = - 0.9 V, $I_t$ = 20 pA, T = 5 K). Unreconstructed unit cells are indicated in green for MoSe$_2$ and dark red for BLG. Approximate moiré pattern unit cells for MoSe$_2$ are outlined in orange. **d**, High resolution STM image of BLG ($V_s$ = - 0.5 V, $I_t$ = 30 pA, T = 5 K).

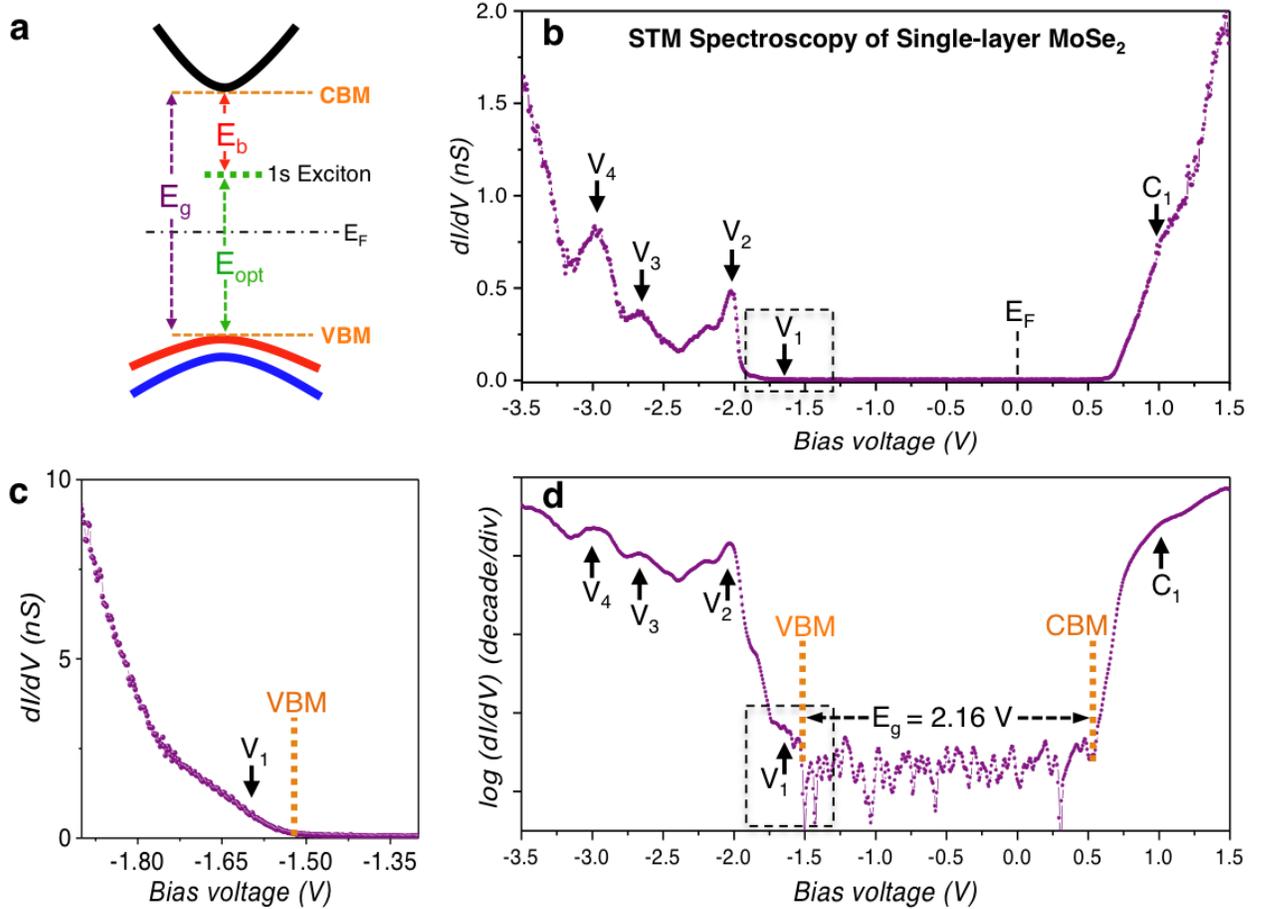

**Figure 2. Electronic structure of monolayer MoSe$_2$ on bilayer graphene (BLG). a**, Energy diagram schematically indicates the electronic bandgap (E$_g$), the optical bandgap (E$_{opt}$), and the excitonic binding energy (E$_b$). **b**, STM dI/dV spectrum acquired on monolayer MoSe$_2$/BLG shows the electronic bandgap and nearby electronic features: V$_{1-4}$ in the valence band (VB) and C$_1$ in the conduction band (CB) (f = 873 Hz, I$_t$ = 5 nA, V$_{rms}$ = 3 mV, T = 5K). **c**, Close-up view of MoSe$_2$ STS (boxed region in **b** and **d**) showing the valence band maximum (VBM) and V$_1$ feature (f = 873 Hz, I$_t$ = 4 nA, V$_{rms}$ = 2 mV, T = 5K). **d**, Logarithm of a typical dI/dV spectrum used in the statistical analysis to obtain E$_g$ (f = 873 Hz, I$_t$ = 5 nA, V$_{rms}$ = 3 mV, T = 5K).

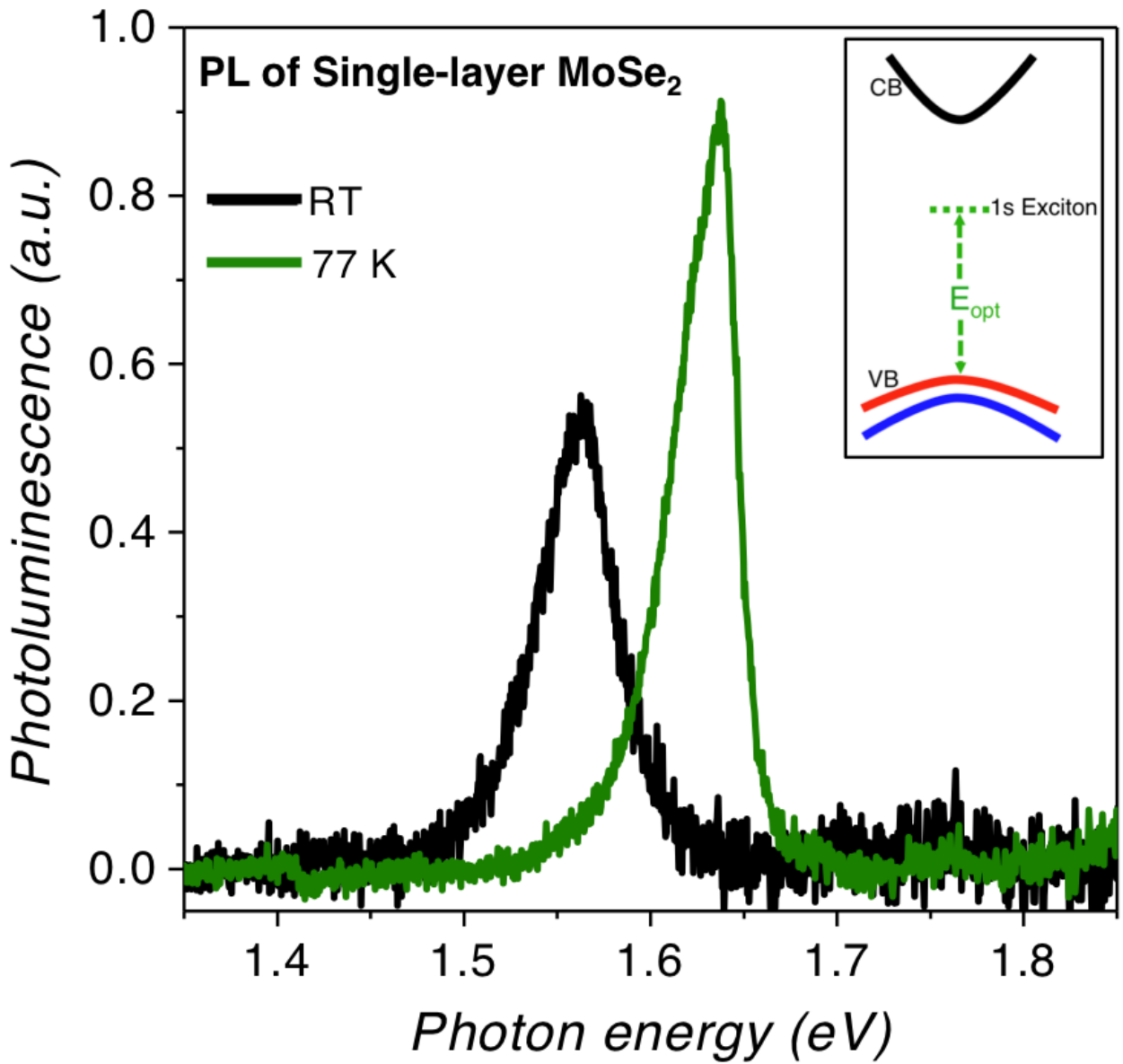

**Figure 3. Optical characterization of monolayer MoSe$_2$ on bilayer graphene (BLG)**: Representative photoluminescence spectra acquired at room temperature (RT) (black) and at 77K (green) for 0.8 ML MoSe$_2$ on BLG/SiC substrate. The photoluminescence at room temperature is centered at 1.55 eV. The peak shifts to 1.63 eV at 77 K. This photoluminescence peak corresponds to the lowest-energy exciton transition (E$_{opt}$) in single-layer MoSe$_2$ (inset).

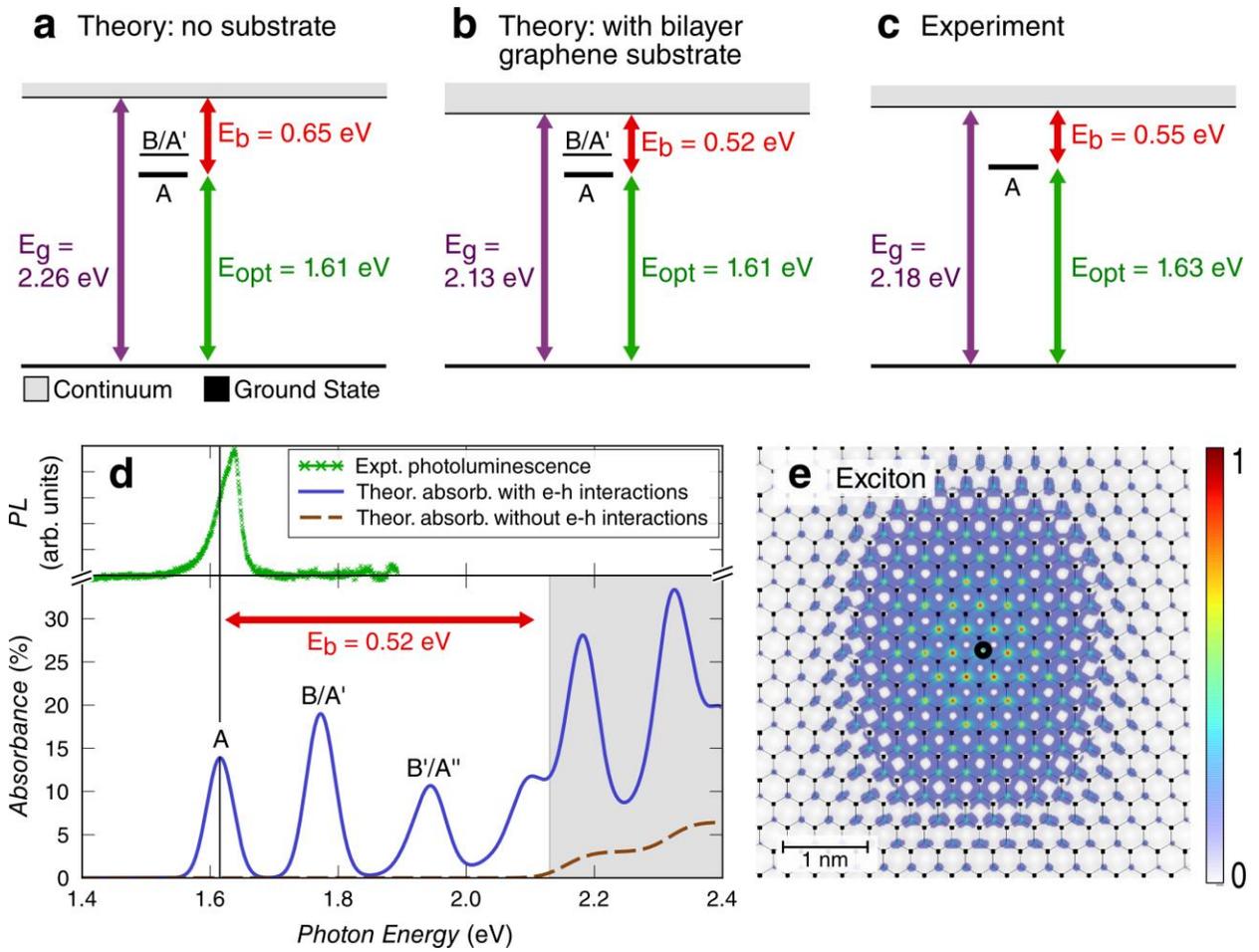

**Figure 4. Comparison between *ab initio* excited-state calculations and single-layer MoSe$_2$ experiment**: Relevant energy levels sketched for **a,** GW-BSE calculation without substrate, **b**, GW-BSE calculation with bilayer graphene (BLG) substrate, and **c,** experimental data. **d**, Calculated optical absorbance of single-layer MoSe$_2$ with and without electron-hole interactions, including BLG screening. A constant Gaussian broadening of $\sigma = 20\ meV$ ($30\ meV$) was used in the curve with (without) electron-hole interactions. The shaded gray area corresponds to energies above the single-particle electronic gap. The experimental PL spectrum measured at 77K is displayed in green. **e,**

Spatial map of the exciton wavefunction corresponding to the excitonic peak labeled 'A' in **a-d** (wavefunction is shown with the hole (black circle) fixed in space). Mo atoms are small black squares, Se atoms not shown.

# Supplementary Materials for

# Observation of giant bandgap renormalization and excitonic effects in a monolayer transition metal dichalcogenide semiconductor

Miguel M. Ugeda[†,*], Aaron J. Bradley[†], Su-Fei Shi[†], Felipe H. da Jornada, Yi Zhang, Diana Y. Qiu, Sung-Kwan Mo, Zahid Hussain, Zhi-Xun Shen, Feng Wang[*], Steven G. Louie, Michael F. Crommie[*].

*Correspondence to: mmugeda@berkeley.edu (MMU), fengwang76@berkeley.edu (FW) and crommie@berkeley.edu (MFC).

† These authors contributed equally to this work.

**This PDF file includes:**

**Materials and Methods**

1. Single-layer MoSe$_2$ growth and structural characterization

*2. Ab-initio* calculations

**Supplementary Text**

1. STM imaging of monolayer MoSe$_2$ on bilayer graphene: interpretation

2. STS reference data on bilayer graphene (BLG) on SiC(0001)

3. MoSe$_2$ electronic bandgap determination

4. Comparison of MoSe$_2$ STS data with ARPES results

5. Tip-induced band bending: tip-height dependent STS

**References**

**Figs. S1 to S7**

## Materials and Methods

### 1. Single-layer MoSe$_2$ growth and structural characterization

Single layers of MoSe$_2$ were grown on bilayer graphene (BLG) substrates in a molecular beam epitaxy (MBE) chamber with base pressure of ~ $2\times10^{-10}$ Torr at the HERS endstation of beamline 10.0.1, Advanced Light Source, Lawrence Berkeley National Laboratory. A reflection high-energy electron diffraction (RHEED) system was used to monitor the growth of MoSe$_2$ layers *in-situ*. BLG substrates were obtained by flash annealing SiC(0001) substrates to ~ 1600 K [30]. Sharp RHEED (Fig. S1A) and low-energy electron diffraction (LEED) patterns (Fig. S1C) of the BLG substrates indicate their high-quality epitaxy and cleanliness. High-purity Mo and Se were evaporated from an electron-beam evaporator and a standard Knudsen cell, respectively. The flux ratio of Mo to Se was controlled to be ~ 1:8. The growth rate was ~ 0.12 ML per minute. During the growth process the substrate temperature was kept at 530 K, and after growth the sample was annealed to 900 K for 30 minutes. Fig. S1B shows the RHEED pattern of a MoSe$_2$ layer with ~ 0.8 ML coverage. A distinct MoSe$_2$ (1×1) pattern can be observed, while the graphene pattern is almost invisible. Fig. S1D shows the co-existing LEED pattern of both the graphene substrate (red circles) and MoSe$_2$ film (green circles). This LEED pattern indicates that MoSe$_2$ layers grow with approximately the same lattice orientation as graphene. However, a more detailed analysis of the diffraction spots of MoSe$_2$ revealed that their intensity was slightly stretched along the rotational direction, indicating that monolayers of MoSe$_2$ also show small rotational misaligned domains ($\theta < \pm 4°$) (red and green dashed lines in Fig. S1D). This is consistent with the moiré patterns observed on the STM images of MoSe$_2$/graphene (see section 1 in the supplementary text). After growth, the sample was transferred directly into the analysis chamber (base pressure of ~ $3\times10^{-11}$ Torr) for low temperature (40 K) core level spectroscopy experiments. The photon energy was set at 70 eV, with energy and angular resolution of 25 meV and 0.1°. Fig. S1E and the zoom-in inset show the characteristic core levels of Mo (36.4 eV of $4p_{3/2}$ orbit and 38.2eV of $4p_{1/2}$ orbit) and Se (54.7 eV of $3d_{5/2}$ orbit and 55.5 eV of $3d_{3/2}$ orbit) of $MoSe_2$. We have also characterized the $MoSe_2$ layers by Raman spectroscopy at room temperature. Raman measurements were performed using continuous wave excitation centered at 488 nm, with the power of 920 μW and a focused spot about 1 μm in diameter. Fig. S1F shows a typical Raman spectrum showing distinctive[31] peaks of $MoSe_2$ at 241.0 cm$^{-1}$ ($A_{1g}$, out of plane) and 287.5 cm$^{-1}$ ($E_{2g}$, in plane) with FWHM (full-width-at-half-maximum) of 5.2 cm$^{-1}$ and 4.0 cm$^{-1}$, respectively.

In order to protect the film from contamination and oxidization during transport through air to the UHV-STM chamber, a Se capping layer with a thickness of ~10 nm was deposited on the sample surface after growth. For subsequent STM experiments, the Se capping layer was removed by annealing the sample at ~ 600 K in UHV for 30 minutes. After this final annealing, the samples were transferred without breaking UHV conditions into the cryogenic STM for surface imaging and spectroscopy. The STM image in Fig. S2 shows the typical morphology of a ~ 0.8ML $MoSe_2$/BLG sample.

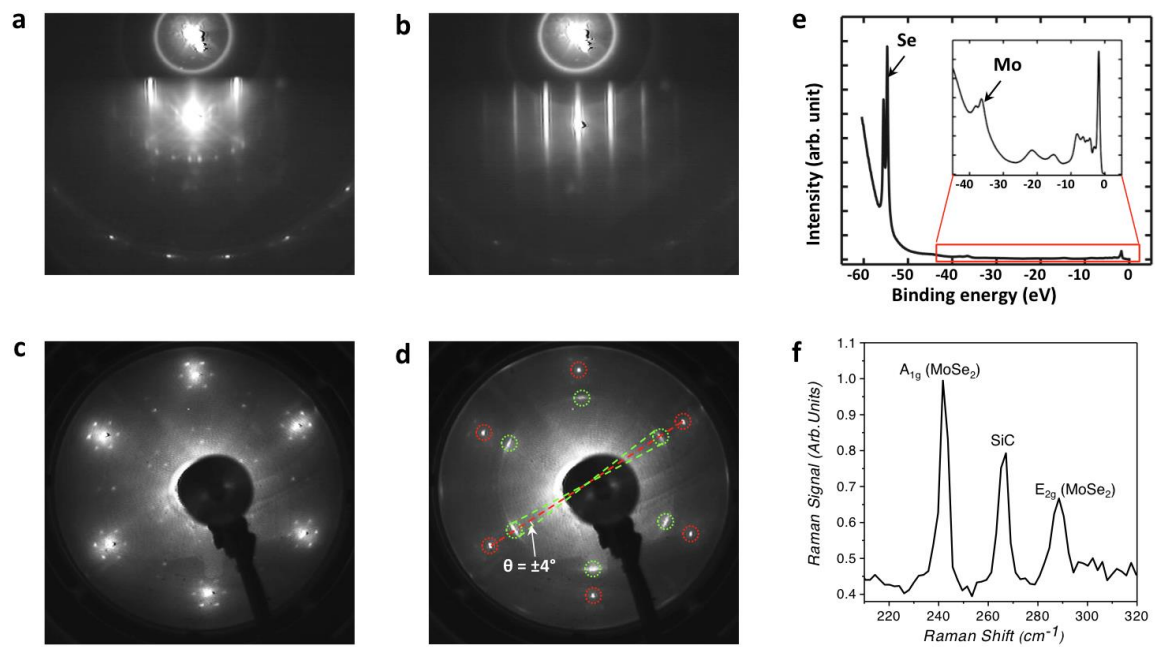

**Figure S1. Characterization of 0.8 ML-MoSe$_2$ grown on bilayer graphene (BLG) on SiC(0001).** RHEED patterns of **a**, BLG and **b**, MoSe$_2$/BLG, respectively. LEED patterns of **c**, BLG and **d**, MoSe$_2$/BLG, respectively. The red and green circles in **d**, indicate the co-existing diffraction spots of both graphene and MoSe$_2$, respectively. The red and green dashed lines in **d**, indicate slight rotational misalignment smaller than ± 4° between MoSe$_2$ and graphene lattices. **e**, Core levels of Se and Mo in MoSe$_2$ film. Inset shows a zoom-in of Mo peaks outlined in the lower red rectangular region in **e**. **f**, Raman spectrum shows two characteristic peaks of single layer MoSe$_2$ at 241.0 cm$^{-1}$ and 287.5 cm$^{-1}$.

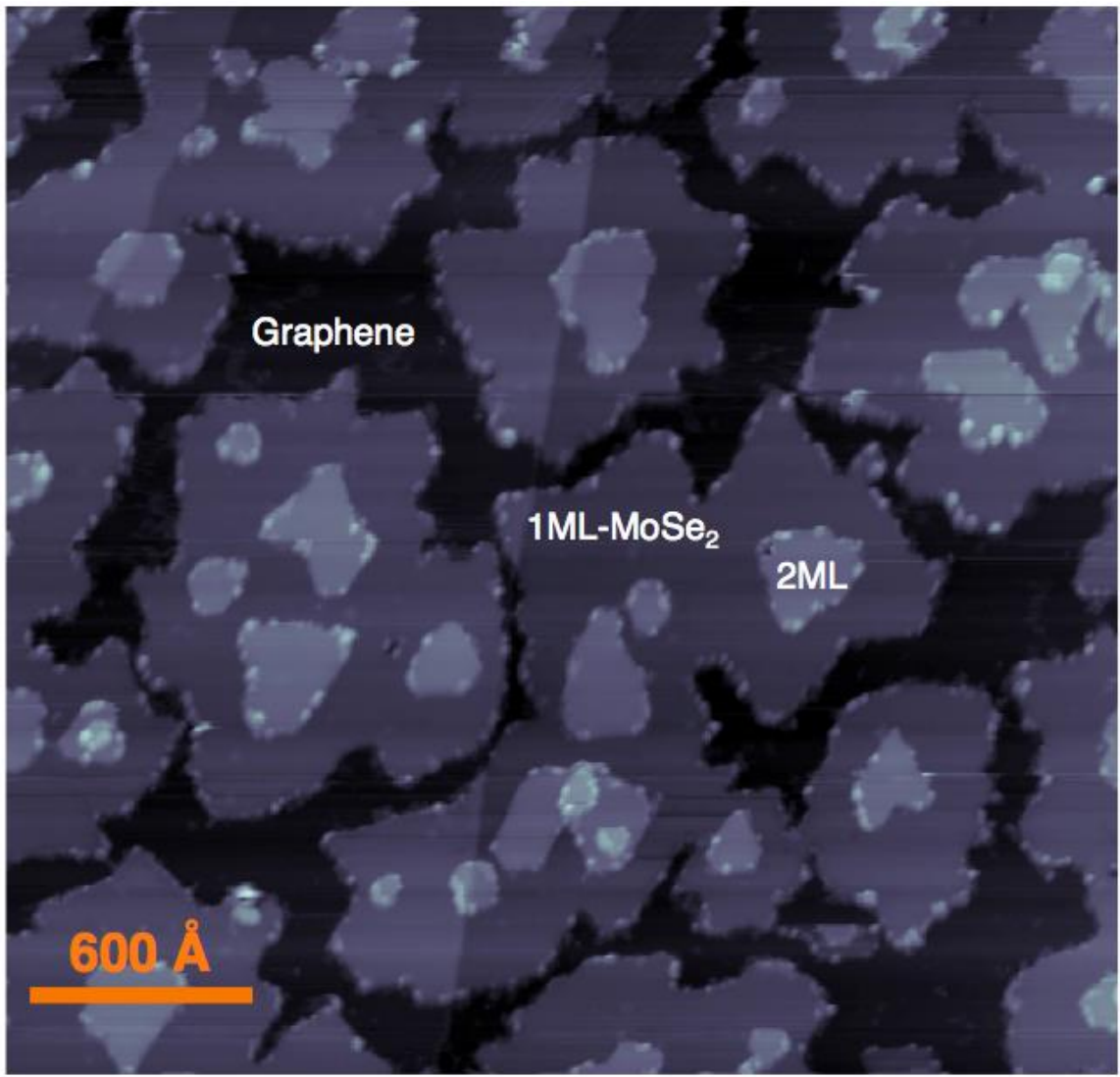

**Figure S2. Morphology of 0.8 ML-MoSe$_2$ grown on bilayer graphene on SiC(0001).** STM image shows large-scale view of the MoSe$_2$/BLG samples studied in this work. Parameters: (3000 Å x 2900 Å, V$_s$ = + 2 .05 V, I$_t$ = 30 pA, T = 5 K).

## 2. *Ab-initio* calculations

We performed our mean-field density functional theory (DFT) calculations in the local density approximation (LDA) using the Quantum Espresso code[32]. The calculations were done in a supercell arrangement with a plane-wave basis using norm-conserving pseudopotentials with a 125 Ry wave function cutoff. We included the Mo semicore 4d, 4p and 4s states as valence states for our DFT, GW and GW-BSE calculations. The distance between repeated supercells in the out-of-plane direction was 25 Å. We fully relaxed the MoSe$_2$ geometry and included spin-orbit interactions as a perturbation as in a previous work[13].

In order to include the substrate screening from the bilayer graphene, we first relaxed a supercell structure with the van der Waals density functional[33] with Cooper exchange[34] and found a separation of 3.4 Å between the MoSe$_2$ layer and the upper graphene layer. We then calculated the irreducible RPA polarizability for the bilayer graphene alone in reciprocal space and mapped the sets of **G** (reciprocal lattice vectors) and **q** wave vectors that describe the polarizability $\chi_{GG'}^{0,bilayer}$ of bilayer graphene to the primitive unit cell of MoSe$_2$. In this process, we neglected local field components in the direction parallel to the substrate, i.e., we set $\tilde{\chi}_{GG'}^{0,bilayer}(\mathbf{q}) = \chi_{GG'}^{0,bilayer}(\mathbf{q})\, \delta_{G_x G'_x} \delta_{G_y G'_y}$. We then calculated a correction to the electronic self-energy of MoSe$_2$ using a correction to the screening,

$$\Delta\varepsilon_{GG'}^{-1}(\mathbf{q}) = \left[1 - v(\mathbf{q}+\mathbf{G})\left(\chi_{GG'}^{0,MoSe2}(\mathbf{q}) + \tilde{\chi}_{GG'}^{0,bilayer}(\mathbf{q})\right)\right]^{-1} - \left[1 - v(\mathbf{q}+\mathbf{G})\chi_{GG'}^{0,MoSe2}(\mathbf{q})\right]^{-1},$$

where $v$ is the truncated Coulomb interaction.

With this expression for $\Delta\varepsilon^{-1}$, we calculated the correction to the self-energy as $\Delta\Sigma = iG(\Delta\varepsilon^{-1})v$, where G is the Green's function. The self-energy of intrinsic MoSe$_2$ was calculated using the GPP model[14], while the correction term was evaluated within the COHSEX approximation.

For the GW calculations, the dielectric matrices were evaluated up to a cutoff of 38 Ry on a 12x12x1 k-grid. Additionally, we calculated a correction to the quasiparticle bandgap as the difference between the gap obtained on a denser k-grid of 21x21x1 and the gap obtained on a coarser k-grid of 12x12x1, where these two calculations where performed with a 15 Ry cutoff for the dielectric matrix. This procedure closes the quasiparticle bandgap calculated without the bilayer graphene substrate by 142 meV, and the gap for the calculation with substrate by 34 meV.

For the Bethe-Salpeter equation (BSE) calculation, we first evaluated the necessary matrix elements on a 21x21x1 k-grid and then interpolated them to a 90x90x1 k-grid, where we diagonalized the BSE Hamiltonian keeping two conduction and one valence band. Further increasing the number of bands to three conduction and three valence bands does not change the absorption spectrum in the energy range from 0 up to 2.4 eV.

The uncertainty in the theoretical calculation arises primarily due to the approximation to the electronic self-energy within the GW approach. Our calculated quasiparticle bandgap is converged to within better than 50 meV, where the error arises primarily due to k-point sampling. However, the GW approximation is known to be accurate to within approximately 100 meV [14]. So, we estimate that our theoretical quasiparticle bandgap should be accurate to within 100 meV.

Similarly, one source of uncertainty when calculating the exciton binding energy is the approximation to the kernel of the electron-hole interaction in the BSE, which is typically a few percent[15]. This uncertainty corresponds to at most ~ 25 meV in $MoSe_2$. On the other hand, we estimate that our numerical solution of the BSE is accurate to within 50 meV due to finite k-point sampling. Therefore, we estimate our exciton binding energy to be accurate to within roughly 50 meV.

# Supplementary Text

## 1. STM imaging of monolayer MoSe$_2$ on bilayer graphene: interpretation

STM images of monolayer MoSe$_2$ having atomic resolution (Figs. 1B, 1C and S3C) show either a honeycomb atomic lattice (Fig. 1B) or, more frequently, a hexagonal lattice (Figs. 1C and S3C). Although the surface of bulk TMDs has been extensively studied with STM in the past, the interpretation of the atomically resolved STM images is not straightforward. Different theoretical approaches suggest that STM images show a hexagonal pattern due to the outermost chalcogen atoms for both polarities[35-38]. However, Altibelli *et al*. showed theoretically that the STM resolution on TMDs ultimately depends on the tip-sample distance[36]. They find that chalcogen atoms are imaged at usual tunneling conditions, in good agreement with other theoretical works, but at small tip-sample distances the contrast is governed by electronic interference effects and the STM images show a honeycomb lattice with contributions from metal atoms and hollow sites. This prediction was corroborated in the past[39] and also is in qualitative agreement with our STM observations.

STM images of monolayer MoSe$_2$ on graphene reveal a superlattice on top of the atomic lattice with periodicity varying from approximately 9.9 Å down to 9.3 Å. This superlattice arises from the distinct lattice parameters of MoSe$_2$ and graphene, as well as their relative stacking orientation (angle θ). As shown in the sketch in Fig. S3A, superposition of equally oriented (θ = 0°) MoSe$_2$ and graphene atomic lattices yields a quasi-commensurate superstructure (a moiré pattern), which is formed due to nearly perfect match between three unit cells of MoSe$_2$ (9.87 Å x 9.87 Å) and four unit cells of graphene (9.84 Å x 9.84 Å). In this case the angle α between the MoSe$_2$ lattice and the moiré periodicity is α = 0°. When the atomic lattices become misaligned and the rotation angle θ increases, the resulting incommensurate moiré pattern shows a decreasing periodicity and α increases (Fig. S3B). This behavior is consistent with our STM observations.

The STM image in Fig. S3C shows a misoriented MoSe$_2$ domain with a moiré periodicity of 9.65 Å x 9.65 Å that is rotated at an angle α = 6° with respect to the MoSe$_2$ atomic lattice. These values correspond to a rotational misalignment of monolayer MoSe$_2$ close to θ = 1.5° with respect to the graphene lattice (see sketch shown in Fig. S3B). In Fig. S3D we plot the periodicity of a collection of moiré patterns measured as a function of the α angle. We have observed regions with α is as large as 16°, which implies rotational misorientations of nearly 4° between both atomic lattices. This is in good agreement with LEED patterns measured for our MoSe$_2$/BLG samples (see Fig. S1D), which indicate that monolayer MoSe$_2$ not only grows aligned with the atomic registry of the graphene substrate (θ = 0°), but also shows other misoriented domains with angles θ < 4°.

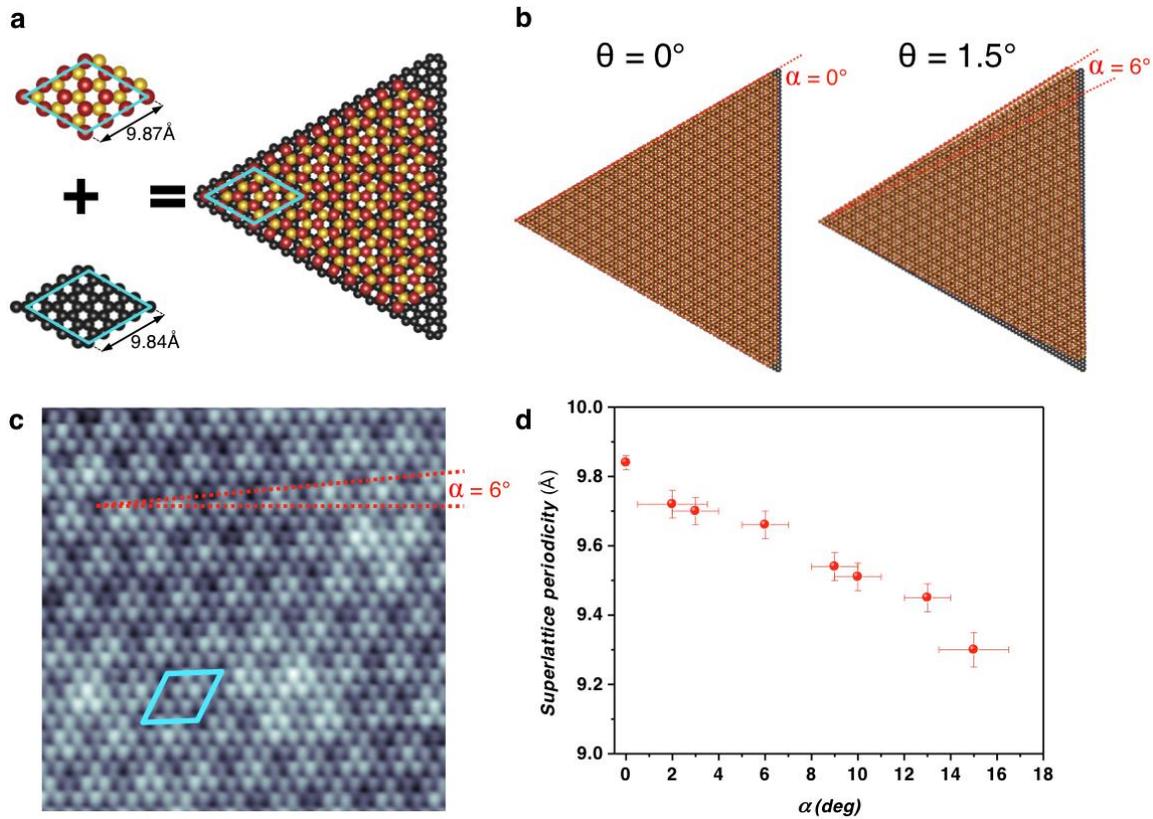

**Figure S3. Moiré pattern formation for MoSe$_2$ on bilayer graphene. a**, Schematic representation of the formation of the quasi-commensurate superperiodicity (i.e., moiré pattern) due to the atomic lattice mismatch between MoSe$_2$ and graphene (3:4). **b**, Example of a small rotational misalignment (θ = 1.5°) between MoSe$_2$ and graphene showing the large effect on α. **c**, Atomic-resolution STM image of monolayer MoSe$_2$ showing an incommensurate 9.65 Å x 9.65 Å moiré pattern with an angle of α = 6°. The rotational misorientation between MoSe$_2$ and graphene lattices is θ = 1.5° (75 Å x 75 Å, V$_s$ = + 0.5 V, I$_t$ = 40 pA, T = 5 K). **d**, Moiré superlattice periodicity plotted as a function of the measured angle α between the atomic lattice vectors and the moiré unit vectors for different regions and samples.

## 2. STS reference data on bilayer graphene (BLG) on SiC(0001)

We utilized a tip calibration procedure during our STS measurements on MoSe$_2$ to avoid tip-induced artifacts in the data. The tip was calibrated for each set of data by first measuring the electronic structure of the BLG on SiC(0001), which has been well characterized in the past[27, 28, [40-42]. The calibration was performed before and after each data acquisition on MoSe$_2$, resulting in tips having appropriate work functions and appropriate BLG spectra. This ensured a good vacuum gap between the tip and sample, and artifact-free MoSe$_2$ spectra.

The low-energy band structure of BLG on SiC(0001) is dominated by the π and π* bands of graphene and exhibits n-type doping. The interlayer coupling between graphene layers and the AB stacking substantially modify the topology of the BLG π and π* bands, sketched in the inset of Fig. S4. While the interlayer coupling causes a split of ∼ 0.4 eV in the π and π* bands[40], the AB stacking breaks the lattice symmetry resulting in the formation of a bandgap opening around the Dirac point[40, 41]. This band structure explains the main features observed in STS experiments. Fig. S4 shows a characteristic reference dI/dV curve acquired on BLG on SiC(0001). The main feature is a dip of width ∼ 150 mV centered at - 390 mV, which corresponds to the bandgap opening around the Dirac point ($E_D$). The asymmetric U shape of the curve, showing the minimum at $E_F$, is due to the asymmetry of the structure of the π* bands around this energy. This spectrum is very similar to other STS spectra reported previously for BLG on SiC(0001)[27, 28, 42].

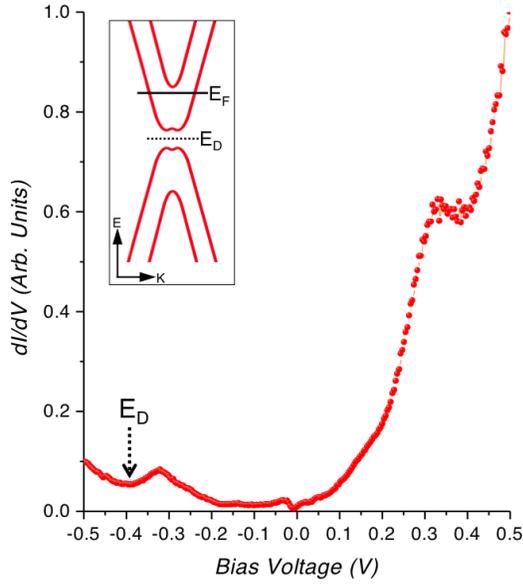

**Figure S4. Tip calibration for STS on bilayer graphene on SiC(0001).** Typical dI/dV spectrum taken on BLG/SiC(0001). The arrow indicates the position of the Dirac point ($E_D$). Parameters: f = 873 Hz, , $I_t$ = 0.5 nA and $V_{rms}$ = 2.8 mV. Inset: Sketch of the electronic band structure of the BLG around the K point.

## 3. MoSe$_2$ electronic bandgap determination

The electronic bandgap value of $E_g$ = 2.18 ± 0.04 eV obtained for monolayer MoSe$_2$ was determined through a statistical analysis of 57 individual STS curves obtained on multiple MoSe$_2$/BLG samples with multiple STM tips. These STS curves were all acquired far from the edges of monolayer MoSe$_2$ patches (d > 1nm). Due to the wide variation in signal strength near a band edge, it was necessary to use the logarithm of the dI/dV curves for accurate gap determination. For each curve, the same procedure was utilized, as follows: we first vertically offset all the data by a factor of 1.1 times the absolute value of the overall minimum value of the spectrum in order to make sure that we could take the logarithm without running into any negative numbers (we tried varying the amount of the offset between 1.1 and 10 times the minimum value and observed no significant change in the final gap value). We then calculated the mean value of the signal within the bandgap to determine the average floor of the spectrum ($C_{g,av}$), as well as the standard deviation ($\sigma$) for signal fluctuations around $C_{g,av}$. After taking the logarithm of the curve, we determined the energies at which the CB and VB edges in the dI/dV signal approached to within $2\sigma$ of $C_{g,av}$. These energies are labeled $E_{VB,2\sigma}$ and $E_{CB,2\sigma}$ (see Fig. S5). Linear fits were then made to the log(dI/dV) spectrum for energies $E_{VB,2\sigma} - \Delta E < E < E_{VB,2\sigma}$ (the VB fit) and for energies $E_{CB,2\sigma} < E < E_{CB,2\sigma} + \Delta E$ (the CB fit). $\Delta E = 150$ mV was determined as the largest energy range over which the linear fit of to the log(dI/dV) signal yielded an $R^2$ value (the coefficient of determination, which indicates how well data is fit by a regression[43]) greater than 0.95 (we find that any value of $\Delta E$ in the range 100 mV $< \Delta E <$ 300 mV results in a mean bandgap well within our error bars). The bandgap edges ($E_{VBM}$ and $E_{CBM}$) were then determined as the points where the VB and CB linear fit lines intersect the bandgap floor determined by $C_{g,av}$. Our reported value of $E_g = 2.18 \pm 0.04$ eV is the average value of $E_{CBM} - E_{VBM}$ determined in this way for the 57 STS curves. The uncertainty was determined by adding the standard errors (i.e., standard deviations of the mean) of the VBM and CBM in quadrature, as well as including the effects of band bending (see sections 4 and 5).

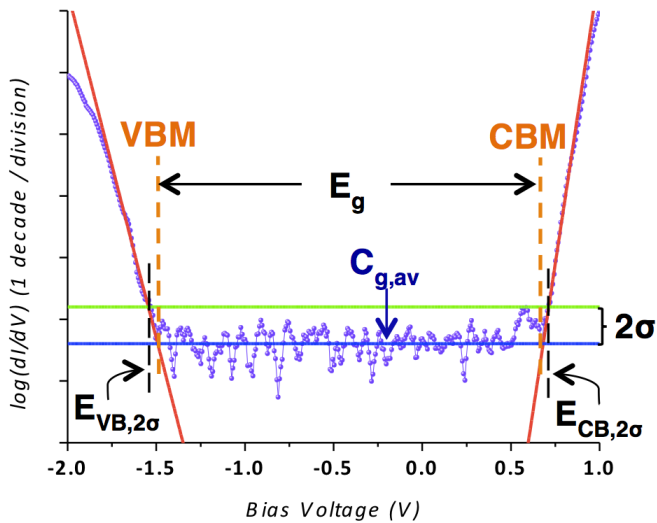

**Figure S5. Bandgap determination.** The bandgap of monolayer MoSe$_2$ was determined from the logarithm of 57 individual STS curves. One such curve is shown in purple. The logarithm of the mean of the signal within the bandgap is depicted in blue as $C_{g,av}$. The green line is two standard deviations above $C_{g,av}$. The red lines on either side of the curve are linear fit lines to the band edges. The gap edges are defined as the energies at which the linear fit lines cross $C_{g,av}$.

## 4. Comparison of MoSe$_2$ STS data with ARPES results

The spectroscopic features $V_{1-4}$ observed in our STS experiments on MoSe$_2$/BLG (Fig S6A) can be interpreted with the aid of the VB electronic structure revealed by ARPES measurements performed previously by some of us[9]. Fig. S6B shows a typical ARPES spectrum of a ~ 0.8 ML MoSe$_2$ sample grown on BLG on SiC(0001). The VB energy region of monolayer MoSe$_2$ is dominated by three dispersive bands with hybridized Mo-*d* and Se-*p* character (bands labeled in Fig S6B as 1-3). In addition to the characteristic MoSe$_2$ electronic features, linear bands from graphene are also visible in the spectrum and mostly disperse beyond the MoSe$_2$ K-point. The VBM of single-layer MoSe$_2$ is located via ARPES at the onset of the split band 1 (-1.53 eV relative to $E_F$) at the K-point, which unambiguously characterizes[9] a single layer of MoSe$_2$. Similarly, STS experiments show the VBM at − 1.55 ± 0.03 eV, which is the onset of the STS feature $V_1$ (Fig. 2C and S6A). Therefore, we can identify the feature labeled $V_1$ with the spin-split band that occurs at the K point of 2D semiconducting TMDs.

The other main STS features in the VB energy range ($V_{2-4}$) are located 0.44 eV ($V_2$), 1.10 eV ($V_3$) and 1.42 eV ($V_4$) below the VBM. A simple comparison with the ARPES spectrum enables us to identify these features with dispersionless regions of bands 1-3 close to the Γ symmetry point. Dispersionless regions of electronic bands lead to van Hove singularities in the density of states which are often easily detectable via STS. Bands 1 and 3 show no dispersion at the Γ point (0.39 eV and 1.08 eV, respectively, below the VBM) (Fig. S6B). This is in good agreement with the position of the STS peaks $V_2$ and $V_4$. The feature $V_3$, which is more broadened than $V_2$ and $V_4$, corresponds to the slightly dispersive band 2 along ΓK. This band shows two dispersionless regions at 1.08 eV and 1.20 eV below the VBM, which also is in good agreement with the STS $V_3$ peak maximum at 1.10 eV below VBM. The much larger conductivity of peaks $V_{2-4}$ compared to peak $V_1$ is due to k-parallel dependence of the tunneling probability (larger k-values lead to lower tunneling probability due to reduced extension of the wavefunction into the vacuum).

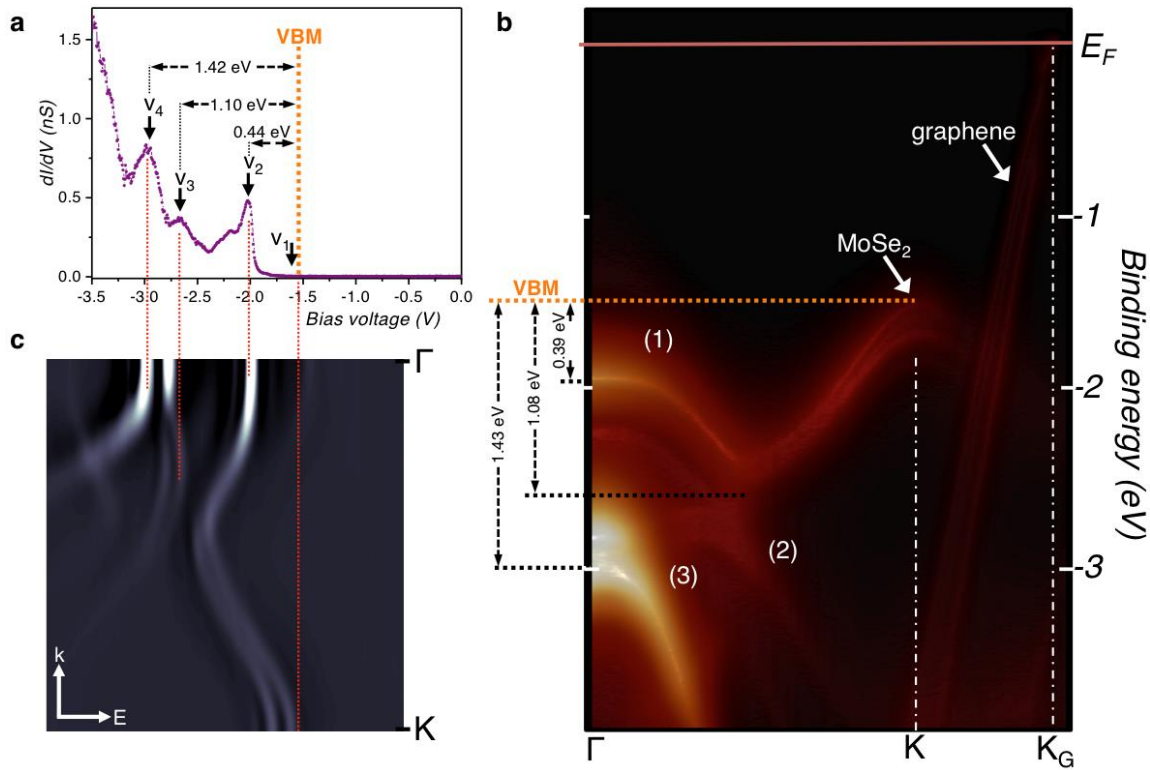

**Figure S6. Interpretation of STS features in the valence band. a**, Occupied states portion of the STS curve shown in Fig. 2B. The energy separations of the $V_{2-4}$ features with respect to the VBM are indicated. **b**, ARPES spectra taken at T = 40K on a submonolayer MoSe$_2$/BLG grown on SiC(0001). The photon energy was set at 70 eV,

with energy and angular resolution of 25 meV and 0.1°. The energy separations of different band onsets with respect to the VBM are indicated. **c**, Second derivative of the ARPES spectrum shown in **b**, compared to STS features in **a**. Red dotted lines project the energy maxima of the $V_{2-4}$ peaks into the ARPES spectrum for a more clear comparison.

## 5. Tip-induced band bending: tip-height dependent STS

Tip-induced band bending (TIBB) is a potential source of error in STS measurements due to unscreened electric fields that can possibly shift the energetic locations of STS features. Here we are able to rule out TIBB as a significant source of error in our measurement of the bandgap of single-layer $MoSe_2$. Although TIBB effects can only be roughly estimated by numerical models[21, 44, 45], they are strongly dependent on accessible experimental parameters such as tip-sample distance (or, equivalently, open-loop tunnel current, $I_t$), tip-sample work function differences, and doping concentration. In order to estimate the influence of TIBB effects on our STS measurements of single-layer $MoSe_2$, we obtained dI/dV curves at a number of different tip-sample distances, as indicated by the initial tunnel current set-point $I_t$ (always obtained using the same initial set-point voltage, $V_s = + 1.5$ V). In fig S7A, we show spectra obtained with a range of initial $I_t$ values covering more than 4 orders of magnitude, equivalent to a variation in tip-sample distance close to 5 Å.

For measurement of TIBB effects on the VBM we focused on the STS feature $V_2$. The reason for this is that the amplitude of feature $V_1$ is too strongly tip-height dependent (i.e., it disappears for larger tip-height distances) since $V_1$ derives from a K-point state in reciprocal space, whereas feature $V_2$ has a more robust tip-height amplitude dependence since it derives from a Γ-point feature (see previous section). Feature $V_2$ is thus the nearest spectroscopic feature to the VBM whose height-dependent energy shifts we can resolve with high accuracy. Since $V_2$ occurs at a higher magnitude of voltage than the VBM, TIBB-induced variation in the location of $V_2$ puts an upper bound on TIBB-induced variation of the VBM. We find that for tunnel current set points ranging over 0.01 nA $\leq I_t \leq$ 500 nA (Fig. S7) the energy variation in the location of $V_2$ is $\Delta V <$ 25 mV. TIBB-induced error in the location of the VBM is thus $\Delta E_{VBM} <$ 25 meV.

In order to deduce the TIBB-induced error in our measured value of the CBM, we directly applied our band-edge finding algorithm (described in the previous section) to determine $E_{CBM}$ in spectra measured for tunnel current set-points ranging over 0.01 nA $\leq I_t \leq$ 500 nA. For this range of tip-sample distances we see no systematic TIBB-induced shift in $E_{CBM}$ and the fluctuations we observe are well within the fluctuation limits that we observed previously. The total TIBB-induced error in our experimental determination of $E_g$ is thus $\Delta E_{g\text{-TIBB}} <$ 25 meV. This TIBB-induced uncertainty is accounted for in the total uncertainty that we report for our measured value of $E_g$ for single layer MoSe$_2$: $E_g = 2.18 \pm 0.04$ eV.

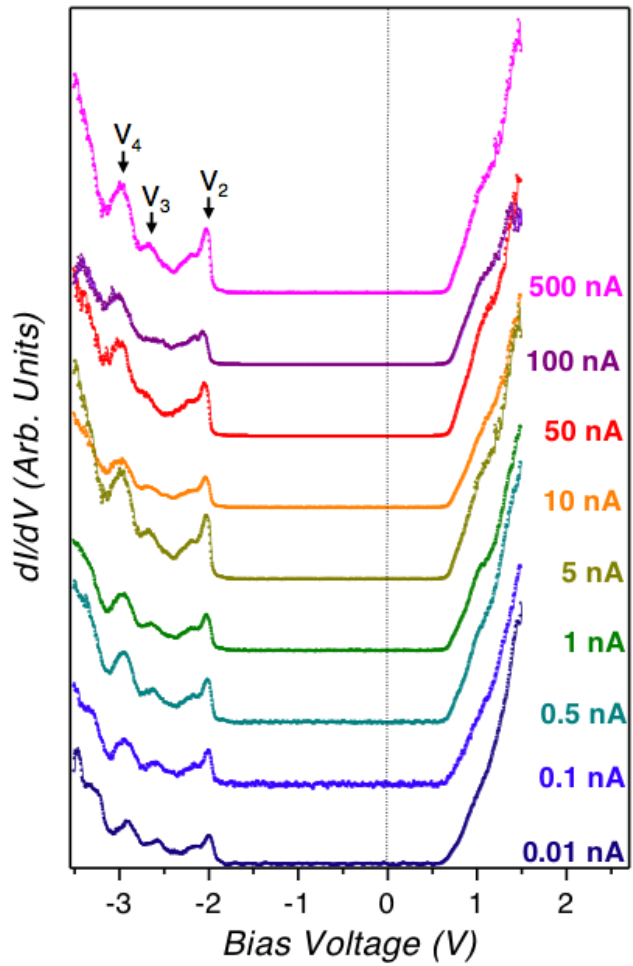

**Figure S7. Tip-induced band bending estimation:** dI/dV spectra acquired at different initial tunneling set-point currents (f = 873 Hz, $V_{rms}$ = 2.8 mV and starting set point voltage = +1.5 V). All curves were taken consecutively with the same tip apex.